\documentclass[amsmath,amssymb,prl,hyperlink,twocolumn]{revtex4}

	\usepackage{graphicx}
	\usepackage{soul}
	\usepackage[colorlinks=true,citecolor=blue,linkcolor=magenta]{hyperref}
	\usepackage[usenames]{color}
	\usepackage{amsfonts}
	\usepackage{color}
	\usepackage{booktabs}
	\usepackage{multirow}
	\usepackage{float}
	
	\def\ket#1{\left\lvert {#1} \right\rangle}

\begin{document}

	\title{Entanglement-Based Machine Learning on a Quantum Computer}
	
	\author{X.-D. Cai,$^{1,2}$ D. Wu,$^{1,2}$ Z.-E. Su,$^{1,2}$ M.-C. Chen,$^{1,2}$ X.-L. Wang,$^{1,2}$ Li Li,$^{1,2}$ N.-L. Liu,$^{1,2}$ \\ C.-Y. Lu,$^{1,2}$ and J.-W. Pan,$^{1,2}$ \vspace{0.2cm}}
	
\affiliation{$^1$Hefei National Laboratory for Physical Sciences at Microscale and Department of Modern Physics, University of Science and Technology of China, Hefei, Anhui 230026, China}
\affiliation{$^2$CAS Centre for Excellence and Synergetic Innovation Centre in Quantum Information and Quantum Physics, University of Science and Technology of China, Hefei, Anhui 230026, China}
	
	
	\begin{abstract}
Machine learning, a branch of artificial intelligence, learns from previous experience to optimize performance, which is ubiquitous in various fields such as computer sciences, financial analysis, robotics, and bioinformatics. A challenge is that machine learning with the rapidly growing ``big data'' could become intractable for classical computers. Recently, quantum machine learning algorithms [Lloyd, Mohseni, and Rebentrost, arXiv.1307.0411] were proposed which could offer an exponential speedup over classical algorithms. Here, we report the first experimental entanglement-based classification of 2-, 4-, and 8-dimensional vectors to different clusters using a small-scale photonic quantum computer, which are then used to implement supervised and unsupervised machine learning. The results demonstrate the working principle of using quantum computers to manipulate and classify high-dimensional vectors, the core mathematical routine in machine learning. The method can in principle be scaled to larger number of qubits, and may provide a new route to accelerate machine learning.
	\end{abstract}
	
	\pacs{}

	\maketitle

There are two main types of machine learning tasks \cite{Book}, namely supervised and unsupervised machine learning. In supervised machine learning, the learner is provided a set of training examples with features presented in the form of high-dimensional vectors and with corresponding labels to mark its category. The aim is to classify new examples based on these training sets. A simple example is a spam filter that sorts incoming emails into spam and non-spam messages by comparing the new emails with old emails already labelled by human. In unsupervised machine learning, the system aims to classify the data into different groups without prior information. An example of unsupervised machine learning is to recognize the object from a landscape background, i.e., to classify the pixels of the image into two groups --- the object and the background. The core mathematical task for both supervised and unsupervised machine learning algorithm is evaluating the distance and inner products between the high-dimensional vectors to analyze the similarity between vectors, which requires a time proportional to the size of the vectors on classical computers. With rapidly growing data size in the modern world, such a task could pose a challenge even for the latest supercomputers.

Recently, it has been shown by Lloyd, Mohseni, and Rebentrost \cite{ML1} that quantum computers, which are naturally good at manipulating vectors and matrices, could provide an asymptotically exponential speed-up over their classical counterparts in performing some machine learning tasks involving large vectors. Consider the task of assigning $N$-dimensional vectors to one of $k$ clusters, each with $M$ representative samples, a quantum computer takes time $O(\mathrm{log}(MN))$. The exponential speed-up of the quantum machine learning algorithm, and its potential wide applications, may make it one of the promising applications of quantum computers \cite{ML1,ML2,ML3}, in addition to Shor's factoring algorithm \cite{shor1,shor2,shor3,shor4,shor5}, quantum simulation \cite{simulation1,simulation2,simulation3,simulation4,simulation5}, and the quantum algorithm for solving linear equation systems \cite{LE1,LE2}.

\begin{figure*}[tb]
        \includegraphics[width=0.7\textwidth]{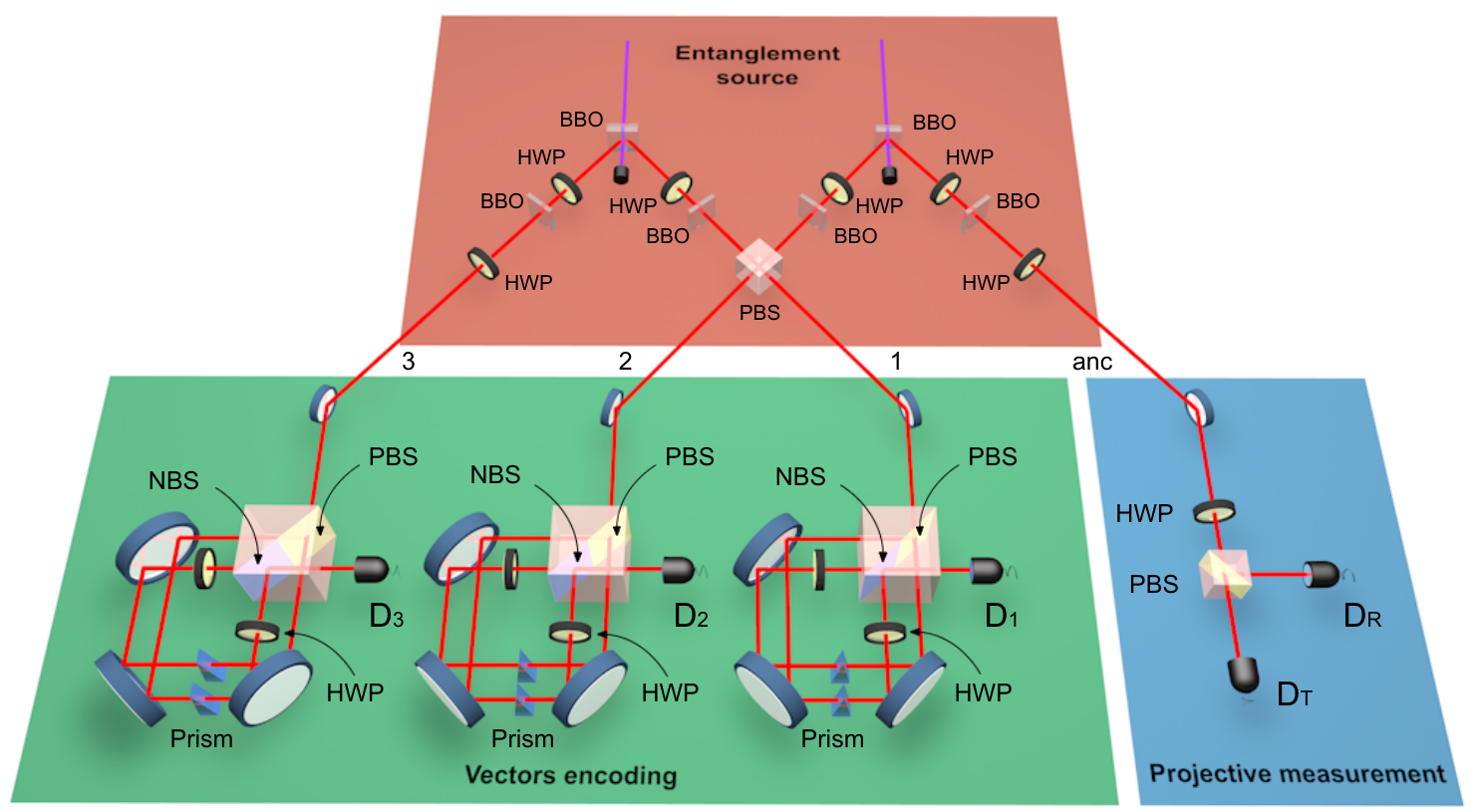}
\caption{Experimental setup for quantum machine learning with photonic qubits. Ultraviolet laser pulses with a central wavelength of 394 nm, pulse duration of 120 fs, and a repetition rate of 76 MHz pass through two type-II $\beta$-barium borate (BBO) crystals with a thickness of 2 mm to produce two entangled photon pairs. The photons pass through pairs of birefringent compensators consisting of a 1-mm BBO crystal and a HWP to compensate the walk-off between horizontal and vertical polarization, and are prepared in the quantum state: $(\ket{H}\ket{V}+\ket{V}\ket{H})/\sqrt{2}$. Two extra HWPs placed in arm 3 and anc are used to transform the state into $(\ket{H}\ket{H}+\ket{V}\ket{V})/\sqrt{2}$. Two single photons, one from each pair, are temporally and spatially superposed on a PBS to generate a four-photon entangled state:  $(\ket{H}\ket{H}\ket{H}\ket{H}+\ket{V}\ket{V}\ket{V}\ket{V})/\sqrt{2}$. The photons 1, 2, and 3 are sent to Sagnac-like interferometers, where each single photon splits into two spatial modes by the PBS with regard to its polarization, and recombines on a non-polarizing beam splitter (NBS). Various vectors are independently encoded into the two spatial modes using HWPs. The specially designed beam splitter cube is half-PBS coated and half-NBS coated. High-precision small-angle prisms are inserted for fine adjustments of the relative delay of the two different paths.  The photons are detected by five single-photon detectors (quantum efficiency $>60\%$), and the two four-photon coincidence events, $D_3 D_2 D_1 D_T$ and $D_3 D_2 D_1 D_R$, are simultaneously registered by a homemade FPGA-based coincidence unit.}
\label{fig1}
\end{figure*}

In this Letter, we report proof-of-principle demonstrations of supervised and unsupervised quantum machine learning algorithm \cite{ML1} on a small-scale photonic quantum processor. The core mathematical task is to assign 2-, 4- and 8-dimensional vectors ($N=2, 4, 8$) to two different clusters with one reference vector ($M=1$) in each cluster. The two clusters are labelled as $A$ and $B$, each with one reference sample vector $\vec{v}_A$ and $\vec{v}_B$, respectively. To classify the new sample which is represented by the vector $\vec{u}$ , one common method is to calculate and compare the distance: $D_A=|\vec{u}-\vec{v}_A|$, and $D_B=|\vec{u}-\vec{v}_B|$. The new sample is assigned to the cluster to which the distance is smaller.

The vectors can be represented with quantum states with a normalization factor, i.e., $\vec{u}=|u|\ket{u}$, $\vec{v}=|v|\ket{v}$. To evaluate the distance $|\vec{u}-\vec{v}|$, a key step in the quantum machine learning algorithm \cite{ML1} is to adjoin an ancillary qubit to the states of the reference and new vectors, creating an entangled state in the form:
	\begin{equation}
		\ket{\varphi}=(\ket{0}_\mathrm{anc}\ket{u}_\mathrm{new}+\ket{1}_\mathrm{anc}\ket{v}_\mathrm{ref})/\sqrt{2}.
	\end{equation}

Next, a single-qubit measurement is made on the ancillary qubit alone (the other qubits are simply ignored), projecting it onto the state:
\begin{equation}
\ket{\phi}=(|u|\ket{0}-|v|\ket{1})/\sqrt{|u|^2+|v|^2}.
\end{equation} The success probability $p$ of this projective measurement can be estimated by repeated measurements. Remarkably, the inner product between $\ket{u}$  and $\ket{v}$ can be directly calculated from the $p$:
\begin{equation}\left\langle u|v \right\rangle=(0.5-p)(|u|^2+|v|^2)/|u||v|,\end{equation} and the distance between $\vec{u}$ and $\vec{v}$ can then be obtained: \begin{equation}D=\sqrt{2p(|u|^2+|v|^2)}.\end{equation} It is important to note that such an estimation can achieve a desired statistical accuracy simply by a sufficient number of repeated measurements, but is \emph{independent} of the size ($N$) of the vectors, which gives a quantum speedup.

This algorithm can be understood intuitively; the more different between the pure states $\ket{u}$ and $\ket{v}$ , the more entangled the equation $(1)$ is. For examples, if  $\ket{u}$ and $\ket{v}$ are identical, then the ancillary qubit is in the state $(\ket{0}+\ket{1})/\sqrt{2}$, separable from the vector qubits, and $p=0$, $D=0$. If $\ket{u}$ and $\ket{v}$ are orthogonal, then the equation $(1)$ is maximally entangled, and $p=0.5$, $D=\sqrt{|u|^2+|v|^2}$.

In our experiment, we use single photons as qubits, where $\ket{0}$ and $\ket{1}$ are encoded with the photon's horizontal ($H$) and vertical ($V$) polarization, respectively. A schematic drawing of the experimental setup is illustrated in Fig.$\,$1. Polarization-entangled photon pairs are generated by spontaneous parametric down-conversion \cite{Kwiat1995} and prepared in the state: \begin{equation}(\ket{0}_{\mathrm{anc}}\ket{0}_{\mathrm{vec}}+\ket{1}_{\mathrm{anc}}\ket{1}_{\mathrm{vec}})/\sqrt{2}.\end{equation}  One photon (anc) is used as the ancillary qubit, and the other one (vec) will be used to encode the reference and incoming vectors using Sagnac-like interferometers (see Fig. 1).

\begin{figure}[tb]
    \centering
        \includegraphics[width=0.48\textwidth]{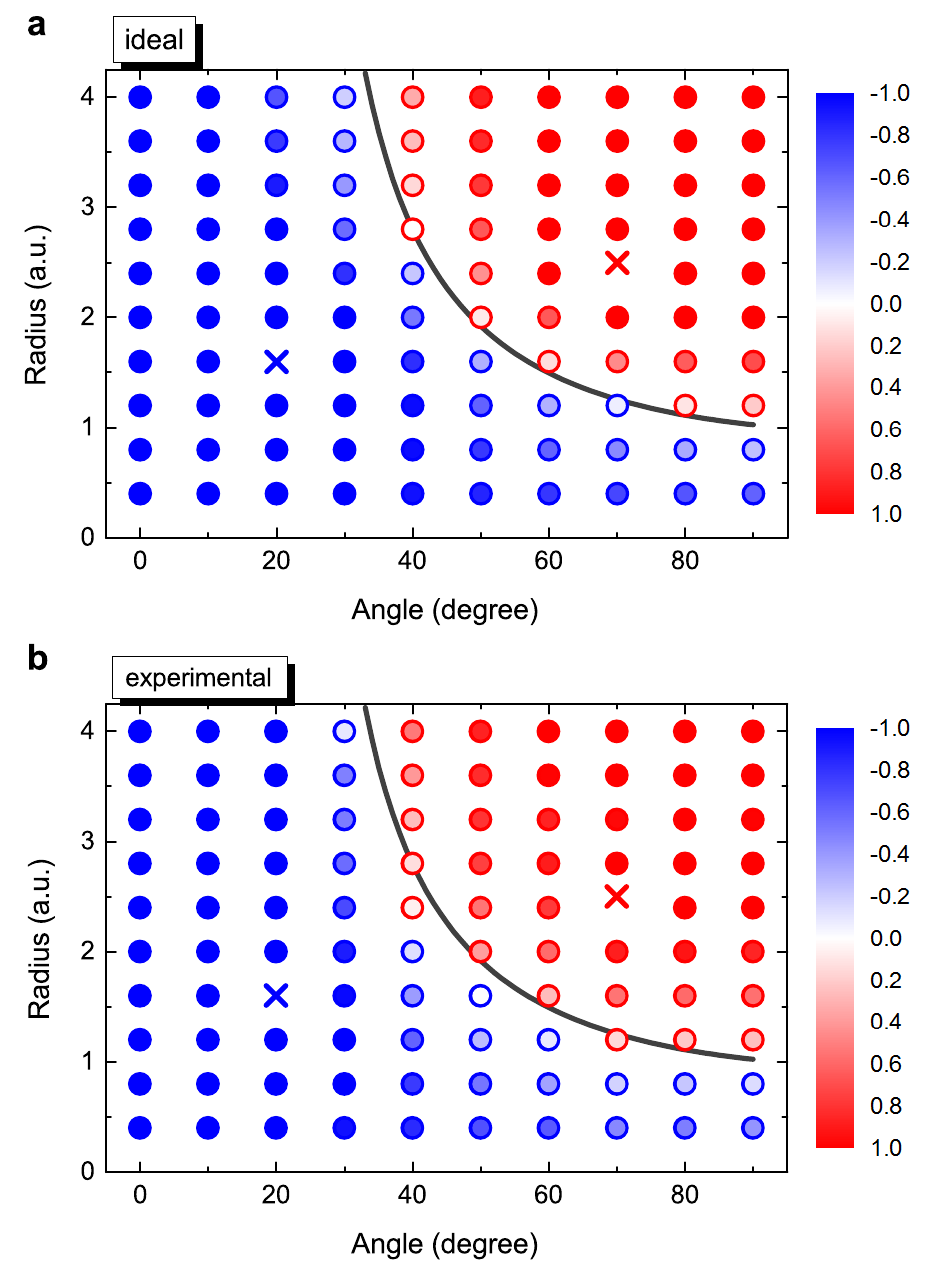}
\caption{Theoretical prediction (a) and experimental results (b) for classifying two-dimensional vectors into two clusters. The red and blue cross are reference vectors. The evaluated value of difference of the the distances of each tested vector to the two reference vectors is coded as the fill color. The result of classification is coded as the edge color (Blue=A, Red=B). The gray line is where the distances are the same in theory. The data acquisition time is 1 second for each vector, collecting about $10,000$ events. The statistical standard deviation is much smaller than the error caused by the imperfection of the entanglement state, thus error bars are omitted. The data is represented in polar coordinate.
}
\label{fig2}
\end{figure}

To generate three- and four-photon entanglement resource state, we create two entangled photon pairs. Two single photons, one from each pair, are temporally and spatially superposed on a polarising beam splitter (PBS). We select the events where one and only one single photon emits from each output. It can be concluded that the four photons are either all $H$ polarized or $V$ polarized, two cases that are quantum mechanically indistinguishable when all the other degrees of freedom of the photons are erased (see the caption of Fig.$\,$1), thus projecting the four photons into the Greenberger-Horne-Zeilinger entangled state \cite{Pan-RMP}: \begin{equation}(\ket{0}_{\mathrm{anc}}\ket{000}_{\mathrm{vec}}+\ket{1}_{\mathrm{anc}}\ket{111}_{\mathrm{vec}})/\sqrt{2}.\end{equation} By projecting one of the four photons into $(\ket{H}+\ket{V})/\sqrt{2}$, we can reduce the four-photon state (6) to three-photon entangled state: \begin{equation}(\ket{0}_{\mathrm{anc}}\ket{00}_{\mathrm{vec}}+\ket{1}_{\mathrm{anc}}\ket{11}_{\mathrm{vec}})/\sqrt{2}.\end{equation} The two-, three-, and four-photon entangled states (5-7) are the entanglement resource used for the classification of the 2-, 4- and 8-dimensional vectors, respectively. We characterize the created multi-photon entangled state using the method of entanglement witness. We obtain the fidelity \cite{fidelity} for the two-, three- and four-photon entangled states to be 0.94, 0.73, and 0.75, respectively, thus prove the presence of genuine multipartite entanglement \cite{toolbox}.

A $2^n$-dimensional vector is encoded with the polarization state of $n$ photonic qubits. For example, a $4$-dimensional vector, (3.42, 1.24, 1.97, 0.72), is represented by the composite quantum state of two single photons with normalization,
\begin{equation}
\begin{split} &\mid{u_2 u_1}\mid\times\ket{u_2 u_1}= \\& 4.2\times(0.866\ket{0}+0.5\ket{1})\otimes(0.94\ket{0}+0.342\ket{1}).\end{split}\end{equation}

\begin{table}[tbp]
\begin{tabular}{cccccc}
\hline
\multicolumn{2}{c}{\multirow{2}{*}{Test Vectors}} & \multicolumn{2}{c}{$D_A-D_B$} & \multirow{2}{*}{Group} & \multirow{2}{*}{Correct?}\\ \cline{3-4}   & & Theory & Exp. & & \\
\hline
1 & (2.00, 0.00, 0.00, 0.00) & -1.45 & -0.93 & A & $\surd$ \\
2 & (0.00, 0.00, 0.00, 2.00) & 0.82 & 0.50 & B & $\surd$ \\
3 & (0.35, 0.20, 0.00, 0.00) & -0.79 & -0.71 & A & $\surd$ \\
4 & (0.23, 0.19, 0.08, 0.07) & -0.54 & -0.51 & A & $\surd$ \\
5 & (1.32, 3.62, 1.57, 4.32) & 0.74 & 0.48 & B & $\surd$ \\
6 & (0.15, 0.17, 0.82, 0.98) & 1.26 & 0.72 & B & $\surd$ \\
7 & (0.18, 0.10, 1.02, 0.59) & 0.98 & 0.76 & B & $\surd$ \\
8 & (0.97, 0.17, 0.17, 0.03) & -1.37 & -0.93 & A & $\surd$ \\
9 & (0.68, 0.25, 0.00, 0.00) & -1.18 & -0.79 & A & $\surd$ \\
10 & (0.83, 0.48, 1.44, 0.83) & 0.67 & 0.17 & B & $\surd$ \\
11 & (1.27, 1.06, 3.48, 2.92) & 1.13 & 0.76 & B & $\surd$ \\
12 & (0.40, 0.40, 0.40, 0.40) & -0.10 & -0.26 & A & $\surd$ \\
13 & (0.09, 0.15, 0.49, 0.85) & 0.80 & 0.55 & B & $\surd$ \\
14 & (0.10, 0.55, 0.06, 0.32) & -0.19 & -0.28 & A & $\surd$ \\
15 & (1.94, 0.34, 0.34, 0.06) & -1.22 & -1.10 & A & $\surd$ \\
16 & (3.42, 1.24, 1.97, 0.72) & -0.34 & -0.39 & A & $\surd$ \\
17 & (0.66, 0.00, 1.80, 0.00) & 0.40 & -0.02 & A & $\times$ \\
\hline
\end{tabular}
\caption{Experimental results for classifying four-dimensional vectors into two clusters. Reference vector A is (1, 0, 0, 0) and B is (0, 0, 1, 1).  The data acquisition time is 2 minutes for each vector, collecting about 500 events.}
\label{table1}
\end{table}

\begin{table*}[tbp]
\begin{tabular}{cccccc}
\hline
\multicolumn{2}{c}{\multirow{2}{*}{Test Vectors}} & \multicolumn{2}{c}{$D_A-D_B$} & \multirow{2}{*}{Group} & \multirow{2}{*}{Correct?}\\ \cline{3-4}   & & Theory & Exp. & & \\
\hline
1 & (2.00, 0.00, 0.00, 0.00, 0.00, 0.00, 0.00, 0.00) & -1.24 & -0.84 & A & $\surd$ \\
2 & (0.00, 0.00, 0.00, 0.00, 0.00, 0.00, 0.00, 0.60) & 0.77 & 0.55 & B & $\surd$ \\
3 & (1.77, 0.00, 0.00, 0.00, 1.24, 0.00, 0.00, 0.00) & -0.92 & -0.52 & A & $\surd$ \\
4 & (0.40, 0.23, 0.11, 0.06, 0.03, 0.02, 0.01, 0.01) & -0.45 & -0.14 & A & $\surd$ \\
5 & (0.00, 0.00, 1.23, 1.23, 0.00, 0.00, 0.33, 0.33) & 0.17 & 0.10 & B & $\surd$ \\
6 & (0.30, 0.03, 0.30, 0.03, 1.12, 0.10, 1.12, 0.10) & -0.11 & -0.24 & A & $\surd$ \\
7 & (0.42, 0.90, 0.35, 0.76, 0.00, 0.00, 0.00, 0.00) & -0.28 & -0.21 & A & $\surd$ \\
8 & (0.54, 0.54, 0.00, 0.00, 0.54, 0.54, 0.00, 0.00) & -0.43 & -0.50 & A & $\surd$ \\
9 & (0.11, 1.24, 0.19, 2.15, 0.06, 0.72, 0.11, 1.24) & 0.40 & -0.17 & A & $\times$ \\
\hline
\end{tabular}
\caption{Experimental results for classifying eight-dimensional vectors into two clusters. Reference vector A is (1, 0, 0, 0, 0, 0, 0, 0) and B is (0, 0, 0, 0, 0, 0, 0, 1). The data acquisition time is 4 minutes for each vector, collecting about 500 events.}
\label{table2}
\end{table*}

To encoded these vectors into the entanglement resource states (5-7), we send the single photons through a PBS where the photon split into two spatial modes according to its polarization. At the two separate spatial modes, controlled unitary operations can be implemented deterministically and independently \cite{Zhou2013}. Thus, we can transform, for instance, the two-photon entangled state (5) into $(\ket{0}\ket{u_1}_{\mathrm{new}}+\ket{1}\ket{v_1}_{\mathrm{ref}})/\sqrt{2}$, where the state $\ket{u_1}$  and $\ket{v_1}$ can be arbitrarily set using wave plates. The two spatial modes are then recombined on a non-polarizing beam splitter. In this way, we create the following 2-, 3- and 4-photon entangled states in the form of Eqn.$\,$(1):
	\begin{equation}
	\begin{split}& \ket{\varphi_2}=(\ket{0}_\mathrm{anc}\ket{u_1}_{\mathrm{new}}+\ket{1}_{\mathrm{\mathrm{anc}}}\ket{v_1}_\mathrm{ref})/\sqrt{2} \\& \ket{\varphi_4}=(\ket{0}_{\mathrm{anc}}\ket{u_2 u_1}_\mathrm{new}+\ket{1}_{\mathrm{anc}}\ket{v_2 v_1}_\mathrm{ref})/\sqrt{2}\\ &\ket{\varphi_8}=(\ket{0}_{\mathrm{anc}}\ket{u_3 u_2 u_1}_\mathrm{new}+\ket{1}_{\mathrm{anc}}\ket{v_3 v_2 v_1}_{\mathrm{ref}})/\sqrt{2} \\
	\end{split}
	\end{equation}
for classifying 2-, 4- and 8-dimensional vectors, respectively.

Figure 2a and 2b display theoretical prediction (ideal) and experimental results of entanglement-based classification of 2-dimensional vectors. The randomly chosen reference vectors are $\vec{v}_A=(1.50,0.55)$ and $\vec{v}_B=(0.86,2.35)$, plotted in polar coordinate as blue and red rectangular cross, respectively. For each new vector $\ket{u_i}_{\mathrm{new}}$, $i=1,2,...100$, two-photon entangled states $\ket{\varphi_2}$ are constructed, and the distance from $\vec{u}_i$ to $\vec{v}_A$ and $\vec{v}_B$ (denoted by $D_A$ and $D_B$) are evaluated from the success probability of the projective measurements on the ancillary photon.

The difference of the distances, $D_A-D_B$,  is color coded in the fill color of each data points in Fig. 2a-b. The sign of $D_A-D_B$  dictates the result of classification: if $D_A-D_B<0$, it is categorized to cluster A (plotted as blue edge color); if $D_A-D_B>0$, it is categorized to cluster  B (plotted as red edge color). The boundary of the two clusters (where $D_A=D_B$) is illustrated as the gray line in Fig. 2a-b. It can be seen that, of the 100 tested samples, two are experimentally misclassified. The misclassification happens for vectors close to the boundary where the absolute error (with an average of $\sim 0.27$), caused by the imperfect two-photon entanglement and dark counts of the single-photon detectors, becomes comparable to $|D_A-D_B|$.

Similar methods can be applied to the classifications of 4- and 8-dimensional vectors based on the construction of three- and four-photon entanglement, with the experimental results listed in Tables I and II, respectively. The precision of distance evaluation is affected by the state fidelity ($\sim$$\,$$75\%$) of the multi-photon entangled state, which is lower compared to that of the two-photon entangled state ($\sim$$\,$$94\%$), mainly caused by double pair emission in parametric down-conversion, the imperfect interference of independent photons on the PBS \cite{Pan-RMP} and the phase fluctuations in the Sagnac interferometers. Among the randomly selected 17 and 9 vectors listed in Tables 1 and 2, respectively, there is one sample misclassified.

\begin{figure*}[tb]
    \centering
        \includegraphics[width=1.00\textwidth]{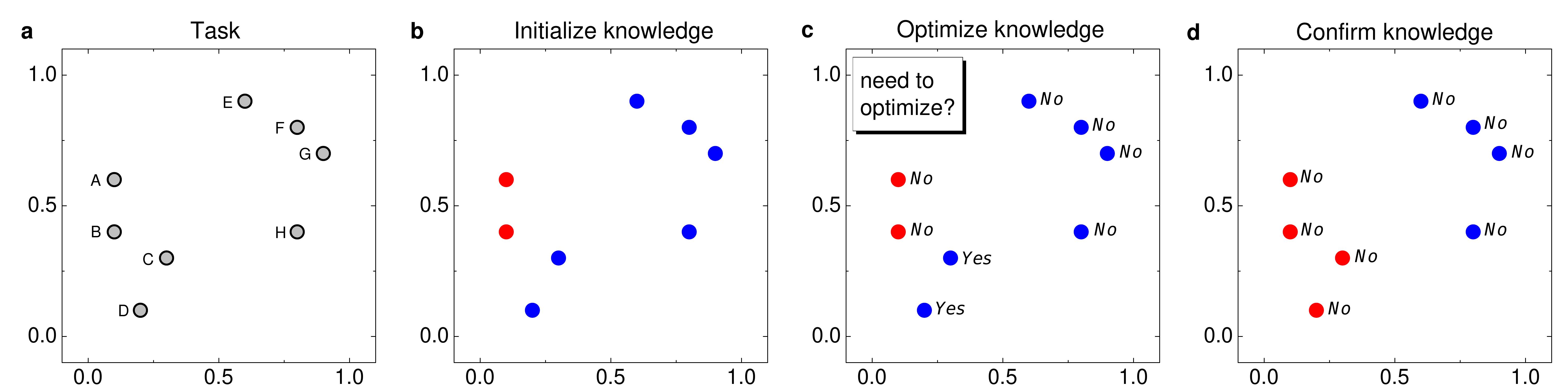}
\caption{Demonstration of unsupervised machine learning. (a) Eight grey circles, labelled from $A$ to $H$, are 2-dimensional vectors to be classified. (b) A random classification is initialized, where $A$ and $B$ belong to red group and $C$, $D$, $E$, $F$, $G$, $H$ belong to blue group. (c) We use the entanglement-based method presented in this paper to experimentally evaluate the distance from each vector to the other vectors within a group. Then the mean distance to both the red and the blue group is calculated: $D_{A-red}=0.12$, $D_{A-blue}=0.67$, $D_{B-red}=0.12$, $D_{B-blue}=0.65$, $D_{C-red}=0.41$, $D_{C-blue}=0.68$, $D_{D-red}=0.33$, $D_{D-blue}=0.58$, $D_{E-red}=0.73$, $D_{E-blue}=0.59$, $D_{F-red}=0.84$, $D_{F-blue}=0.57$, $D_{G-red}=0.91$, $D_{G-blue}=0.57$, $D_{H-red}=0.77$, $D_{H-blue}=0.50$. It can be seen that $C$ and $D$ are closer to red group but were wrongly classified into blue group, whose labels are therefore changed. The labels of the other six vectors remain unchanged. (d) The optimizing process is repeated in the new configuration until there is no change. In this configuration every vector is in the group with a closer mean distance, and the system can confirm the configuration as an ultimate classification result.
}
\label{fig3}
\end{figure*}

The quantum mechanical way of evaluating the vector distance demonstrated above are the core mathematical subroutine for other machine learning tasks, for example, supervised nearest-neighbour algorithm and unsupervised machine learning algorithm. In supervised nearest-neighbor algorithm, each test vector is analyzed by evaluating the distance between itself and all the training vectors, and then categorized into the group of the nearest training vector. When new training vectors are offered, the system will adjust the judgment of classification by analyzing the distances in the new configuration. An example with training sample $M=2$ is shown in Supplemental Fig.$\,$S1.

In unsupervised machine learning, no training vectors are provided, and the system need to realize a reasonable classification by iterating to calculate the distance between different vectors. The algorithm includes three steps: (1) Initialize a random classification. (2) For each vector $v_i$, the learner calculates the distance between $v_i$ and all vectors in a group. The $v_i$ is classified into a group to which the average distance is minimal. (3) Repeat step 2 until no vector needs to change its group.  An example with $M=4$ is demonstrated in Fig.$\,$3.

Note that the current experimental scheme can in principle achieve an exponential speedup with respect to the dimension $N$ of the vectors, but not to the number of training samples $M$. To demonstrate a speedup in numbers of manifold vectors $M$, future studies are planned to design quantum circuits involving $M+1$ level qudits. High-dimensional quantum states can be encoded using, for example,  photons' degree of freedom of orbital angular momentum \cite{OAM}.

In summary, we have performed the first experimental demonstration of machine learning on a photonic quantum computer. Our work demonstrates that the manipulation of high-dimensional vectors and the estimation of the distance and inner product between vectors, a ubiquitous task in machine learning, can be naturally done with quantum computers, thus proved the suitability and potential power of quantum machine learning. The ability of manipulating large vectors---combined with previously realized methods for solving systems of linear equations \cite{LE1,LE2} and Hamiltonian simulation \cite{sparse}---on quantum computers, may provide a useful quantum toolkit for dealing with the ``big data''.

\vspace{0.1cm}
\noindent \textit{Acknowledgement}: This work was
supported by the National Natural Science Foundation
of China, the Chinese Academy of Sciences, and the
National Fundamental Research Program (under Grant
No. 2011CB921300).

\newpage

\section*{Supplemental Material}
\renewcommand{\thefigure}{S\arabic{figure}}
 \setcounter{figure}{0}
\renewcommand{\theequation}{S.\arabic{equation}}
 \setcounter{equation}{0}
 \renewcommand{\thesection}{S.\Roman{section}}
\setcounter{section}{0}
%



\begin{figure}[h]
    \centering
        \includegraphics[width=0.5\textwidth]{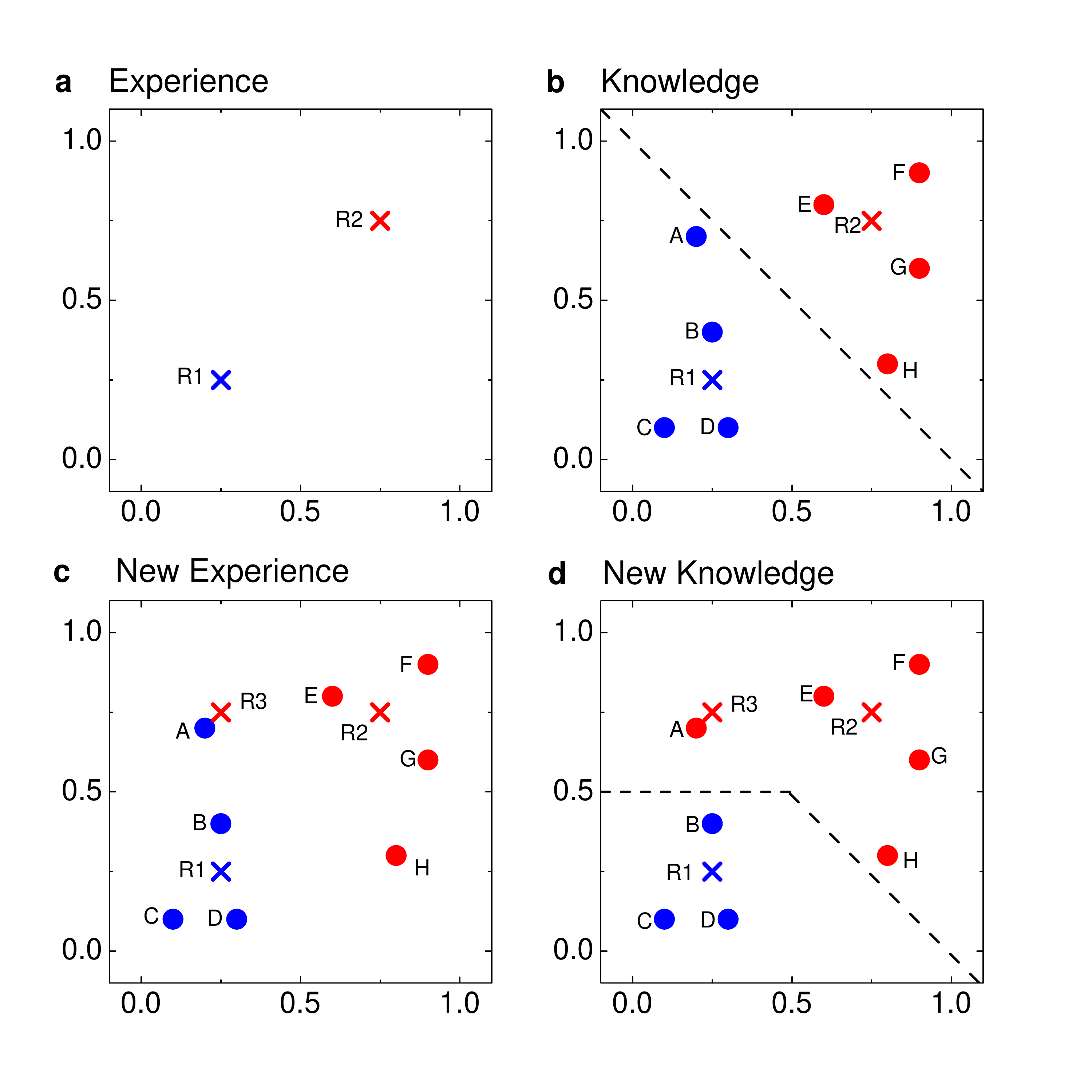}
\caption{Experimental result of supervised nearest-neighbour machine learning algorithm (a) The blue (R1) and red (R2) crosses are two training vectors, representing two different groups (blue and red), respectively. (b) Eight testing vectors are labelled from $A$ to $H$. For each vector, its distance to $R1 $ and $R2$ is experimentally evaluated: $D_{A-R1}=0.54, D_{A-R2}=0.58, D_{B-R1}=0.24, D_{B-R2}=0.62, D_{C-R1}=0.20, D_{C-R2}=0.90, D_{D-R1}=0.21, D_{D-R2}=0.83, D_{E-R1}=0.75, D_{E-R2}=0.54, D_{F-R1}=1.05, D_{F-R2}=0.62, D_{G-R1}=0.86, D_{G-R2}=0.60, D_{H-R1}=0.67, D_{H-R2}= 0.61$. The test vectors are classified into the group with closest distance. Therefore, $A$, $B$, $C$, $D$ belong to the blue group, and $E$, $F$, $G$, $H$ belong to red group. The dash line in the figure represents the theoretical boundary between the two groups. (c) A new training vector $R3$ is provided, and the system needs to improve its classification taking into account this new vector. (d) The distance between each vector $A$ to $H$ to the training vectors $R1$, $R2$, and $R3$ is experimentally evaluated: $D_{A-R1}=0.54, D_{A-R2}=0.58, D_{A-R3}=0.45, D_{B-R1}=0.24, D_{B-R2}=0.62, D_{B-R3}=0.37, D_{C-R1}=0.20, D_{C-R2}=0.90, D_{C-R3}=0.67, D_{D-R1}=0.21, D_{D-R2}=0.83, D_{D-R3}=0.67, D_{E-R1}=0.75, D_{E-R2}=0.54, D_{E-R3}=0.55, D_{F-R1}=1.05, D_{F-R2}=0.62, D_{F-R3}=0.86, D_{G-R1}=0.86, D_{G-R2}=0.60, D_{G-R3}=0.83, D_{H-R1}=0.67, D_{H-R2}= 0.61, D_{H-R3}=0.81$. As a result the label of vector $A$ has to be changed from red to blue, as it is closer to $R3$ than $R1$, while other testing vectors remain unchanged. The boundary in this new configuration is accordingly shifted, represented as two linked dash lines.
}
\label{figS1}
\end{figure}


\begin{thebibliography}{10}
\bibitem{fidelity} The state fidelity is a measure of to what extent the desired state is generated. It can be calculated by the overlap of the experimentally produced state with the ideal one.
\end{thebibliography}
\end{document}